\documentstyle[preprint,aps]{revtex}

\newcommand{\lapprox}{{\renewcommand{\arraystretch}{0.3}
\begin{array}{c} {\tiny < } \cr {\tiny \sim} \end{array}}}
\begin{document}
\bibliographystyle{/usr/lib/texmf/tex/revtex/prsty}
\draft
\renewcommand{\baselinestretch}{1.8} \small \normalsize
\title{
Benchmark Calculations for the Triton Binding Energy for Modern NN Forces and
the $\pi$-$\pi$ Exchange Three-Nucleon Force
}
\author{ A.~Nogga, D.~H\"uber$^{\dagger}$, H. Kamada $^{\ddagger}$ and
  W.~Gl\"ockle\footnote{email: Walter.Gloeckle@hadron.tp2.ruhr-uni-bochum.de} }
\address{
Institut f\"ur theoretische Physik II, Ruhr-Universit\"at Bochum,
D-44780 Bochum, Germany \\
$^{\dagger}$ Los Alamos National Laboratory, 
Theoretical Devision, MS B283, Los
Alamos, NM 87545, USA \\
$^{\ddagger}$ Institut f\"ur Kernphysik, Fachbereich 5, Technische Hochschule
Darmstadt, D-64289 Darmstadt, Germany }
\date{\today}
\maketitle
\narrowtext
\begin{abstract}
\renewcommand{\baselinestretch}{1.1}
\tiny \normalsize
We present high precision benchmark calculations for the triton binding energy
using the most recent, phase equivalent realistic nucleon-nucleon (NN)
potentials and the 
Tuscon-Melbourne $\pi$-$\pi$ three-nucleon force (3NF). That 3NF
is included with partial waves up to a total two-body
angular momentum of $j_{max}=6$. It is
shown that the inclusion of the 3NF slows down the convergence
in the partial waves and $j_{max}=5$ is needed in order to achieve converged
results within a few keV. We adjust the cut-off parameter $\Lambda$ in the form
factors of the 
Tuscon-Melbourne 3NF separately for the different NN potentials to the triton binding energy. This provides
a set of phenomenological three-nucleon Hamiltonians which can be tested in
three-nucleon scattering and systems with  $A>3$.  A connection between the probability to find two nucleons at
short distances in the triton and the effect of that 3NF on the triton
binding energy is pointed out.
\end{abstract}
\begin{list}{}{\setlength{\labelwidth}{3cm} 
                        \setlength{\leftmargin}{5cm}
                        \setlength{\labelsep}{0.4cm}}
\renewcommand{\baselinestretch}{1.1}
\tiny \normalsize
\renewcommand{\makelabel}{PACS numbers:\hfill}
\item 21.30.-x, 21.45.+v, 27.10.+h, 21.10.Dr
\renewcommand{\makelabel}{keywords:\hfill}
\item triton binding energy, three-nucleon forces, \\Faddeev calculations,
two-nucleon correlation functions
\end{list}

\narrowtext
It has been possible to include a three-nucleon force (3NF) into Faddeev
calculations for the 3N bound state since many years
\cite{bomelburg83a}
\nocite{bomelburg83b}
\nocite{ishikawa84}
\nocite{chen85}
-\cite{bomelburg86}. 
In configuration space 
calculations were performed up to a maximal two-body angular momentum of
$j_{max}=4$ \cite{chen86}. The same has been achieved 
using an admixture of configuration and
momentum space 
\cite{sasakawa86}
\nocite{ishikawa86}
-\cite{wu93}. In momentum space no efforts beyond $j_{max}=2$ have been
reported up to now \cite{stadler91} \cite{stadler95}. 

Recently, due to a new way of partial wave decomposition (PWD) for the 3NF in
momentum space \cite{huberapp}, 
it became possible to include higher partial waves
of the 3NF in momentum space with $j>2$. The old PWD \cite{coon81}
\cite{huberphd} 
used up to now leads 
to untractable numerical instabilities for partial waves with $j>2$. Results
published up to now containing the 3NF up to $j_{max}=2$ are however not
affected by that defect in the old PWD. One aim of this study is to extend
the momentum space calculations for the triton binding energy to higher
partial waves to demonstrate convergence within an accuracy of a few keV.

The other aim is provoked by an ambiguity in the Tuscon-Melbourne (TM)
$\pi$-$\pi$ exchange 3NF \cite{coon79} \cite{coon81}. In the
TM 3NF the strong meson nucleon form factors are parametrised in a standard
manner by a certain cut-off parameter $\Lambda$, whose value is only roughly
known. That parameter $\Lambda$ acts like a
strength factor of that 3NF and the 3N binding energy is quite sensitive to
$\Lambda$. A variation of $\Lambda$ within
one pion mass causes differences in the 3N binding energy of about 2 MeV. One
can add additional two-meson exchange 3NFs \cite{coon93} \cite{stadler95} 
like the $\pi$-$\rho$ potential, which counteracts the attraction of the $\pi$-$\pi$
potential. Chiral perturbation theory suggests many more structures
\cite{weinberg90} \cite{ordonez92}. 
The realisation of the multitude of 3NFs in Faddeev
calculations, which technically are still based on PWD, is a highly
non-trivial technical challenge still to be overcome. 
In this situation, where there is not yet a theory which predicts consistent
NN and 3N forces, a phenomenological approach appears to be
justified. Therefore the second aim is to adjust the triton binding energy for
a certain 3NF in conjuction with a NN force.
Such an approach has been already taken by the Urbana-Argonne
collaboration \cite{carlson83} \cite{schiavilla86}. 
Performing now this fit for several modern NN potentials 
one gets a number of 3N
Hamiltonians  which all give the same (correct) triton binding energy. Using
these 
models one can then explore the 3N continuum and search for interesting
3NF effects in elastic nd scattering and in nd breakup observables. Also one
might go on to $A>3$ systems, which has already been pioneered in
\cite{pudliner95}. 

Here in this paper we fit $\Lambda$ for the most recent,
phase-equivalent realistic NN potentials. These are the phenomenological
potential AV18 \cite{av18}, the phenomenological potentials of
the Nijmegen group Nijm~I, Nijm~II and Reid~93 and their meson theoretical
potential Nijm~93 \cite{nijm93}, and the
meson-theoretical CD-Bonn potential \cite{cdbonn}. Note that Nijm~II was
recently refitted in the $^{1}P_{1}$-wave in order to remove an unphysical
bound state in that wave at -964 MeV \cite{stokspriv}. The potentials Nijm~II and
Reid~93 are purely local, whereas Nijm~93, Nijm~I and AV18 carry a small
non-locality in form of $p^{2}$-terms. CD-Bonn, which is defined in momentum
space, is strongly 
non-local and carries all the Dirac structure of the nucleon. All these
potentials are fitted perfectly to the recent Nijmegen phase shift 
analysis \cite{nijmphases} with a $\chi^{2}$ per datum very close to one; only Nijm
93 is fitted less perfectly.

It is clear that these NN forces and that 3NF are inconsistent. This is a
trivial statement for the phenomenological NN forces, but even for the meson
theoretical ones there is no consistent scheme behind the forces. Therefore
fitting $\Lambda$
is just a zeroth order, purely phenomenological step, which will lay some
ground to do exploratory steps in 3N scattering and for systems with $A>3$. This
might be accepted until there are predictive, generally
accepted consistent NN and 3N forces. Going into this direction is the very
promising work of
the Bochum group \cite{ruhrpot} \cite{eden96}. 
Their model should be tested in the future.

The Faddeev equation we are using for the 3N bound state including a 3NF reads
\cite{stadler91} 
\begin{equation}
\label{1}
\left| \psi\right> = G_{0}\ t\ P\ \left| \psi\right>
                   + G_{0}\ (1+t\ G_{0})\ V_{4}^{(1)}\ (1+P)\ 
                   \left| \psi\right>
\end{equation}
Here $G_{0}$ is the free three nucleon propagator, $t$ the two-body $t$-matrix
and $P$ the sum of a cyclic and an anti-cyclic permutation of the three
nucleons. $\left| \psi\right>$ is the Faddeev amplitude, from which one
determines the wave function $\left| \Psi\right>$ via
\begin{equation}
\label{2}
\left| \Psi\right> = (1+P)\ \left| \psi\right>
\end{equation}
We use the fact that
all 3NFs considered up to now can be split into three parts, each of them
being 
symmetric under exchange of two of the three particles:
\begin{equation}
\label{3}
V_{4}=V_{4}^{(1)}+V_{4}^{(2)}+V_{4}^{(3)}
\end{equation}
For example $V_{4}^{(1)}$, occurring in eq.~(\ref{1}), is symmetric under
exchange of particles 2 and 3. 
We solve eq.~(\ref{1}) in momentum space using a 
PWD. For details see \cite{stadler91} \cite{glocklefb}.

Before we show our results let us comment on the numerics. For
the discretisation of the Jacobi momenta $p$ and $q$ (for the notation see
\cite{glocklefb}) we use 40 and 36 Gaussian points, respectively. The cut-offs
of the integrals in $p$ and $q$ are 60 fm$^{-1}$ and 20 fm$^{-1}$,
respectively. For the angular integration introduced by the PWD of the
permutation 
operator $P$ we use 16 points. The nucleon mass is chosen as $m=938.9$ MeV.
Using these sets of grid points we achieve a
numerical accuracy in the binding energy of about 0.1\%. A 
good measure for the numerical accuracy of the wave function is to evaluate
the energy 
expectation value $\left< H\right>\equiv \left< \Psi \right| H \left| \Psi
\right> = \left< \Psi \right| H_{0} \left| \Psi \right> +\left< \Psi \right| V
\left|  \Psi \right> $ and compare it to the energy eigenvalue of eq.~(\ref{1}). We find
that these two numbers differ always by less than
0.05\%. 

First we document in Table \ref{tab2} the convergence of the triton binding
energy with and without 
3NF. For that purpose we chose the AV14 \cite{av14} NN force together with
the TM 
$\pi$-$\pi$ 3NF ($\Lambda=5.13 m_{\pi}$). For that specific model we can
compare to the very recent results of the Pisa group. 
Let us first consider the results as a function of $j_{max}$ without 3NF.
It can be 
seen that in order to reach an accuracy of 0.1\% one has to take
into account all partial waves up to a maximal total two-body angular momentum
of $j_{max}=4$. (For some other potentials we found a somewhat faster
convergence, but with all potentials we considered we reached that accuracy at
$j_{max}=4$.) We also list the expectation values of kinetic energy    
$\left< H_{0}\right> $, NN potential energy $\left< V\right>$, and 3N
potential energy $\left< V_{4}\right>$.  
For the calculations with 3NF we included two 3NFs calculated somewhat
differently as is explained now. In \cite{huberapp} 
we introduced a new way
of PWD for the 3NF. This 
was necessary because our old method for the PWD of the 3NF \cite{coon81}
\cite{huberphd} leads to an untractable numerical problem for partial waves with
$j>2$, as was demonstrated in \cite{huberapp}. (The results, however,  achieved
with the old 
PWD and $j\le 2$ are numerically correct.) The basic idea of the new PWD is, to
split the 3NF into two quasi two-body parts. The PWD of these quasi two-body
parts can be done safely if one chooses a proper basis. This requires several
changes of Jacobi variables, which looks as
\begin{equation}
\label{4}
V_{4}^{(1)}=P_{1\leftrightarrow 2}\ W_{2}\ P_{2\leftrightarrow 3}\ W_{3}\
            P_{3\leftrightarrow 1}
\end{equation}
The three different Jacobi sets are labeled as 1, 2 and 3 and the operators
$P_{i\leftrightarrow j}$ connect basis $i$ with basis
$j$. $W_{2}$ and $W_{3}$ are the two quasi two-body parts of $V_{4}^{(1)}$.
Now it is obvious from the form of eq.~(\ref{4}) that one has to insert four
times the completeness relations for the respective basis 2 and 3 in order to
calculate the
matrix elements of $V_{4}^{(1)}$. We refer to that insertions in the following
as the inner basis of the 3NF. The number of partial waves in that inner basis
is unrestricted, but in practise one can cut it off. That maximal
two-body angular momentum will be called in the following the inner
$j_{max}$ of the 3NF (For more details about all that see \cite{huberapp}). 

We demonstrate the convergence with respect to the
inner $j_{max}$ by showing the results for two
3NFs with an inner $j_{max}$ of 5 and 6. We see from Table
\ref{tab2} that the contribution to the triton binding energy of the inner
partial 
waves with $j=6$ is always less than 0.1\%. Therefore we restrict ourself
in the following always to $j_{max}=5$ for the inner basis of the 3NF. A quick
glance on Table \ref{tab2} also reveals that $\left< H_{0} \right>$ and $\left<
  V\right>$ changes somewhat if a 3NF is included and   
$\left< V_{4} \right>$  is only  about 2 \% of $\left< V \right>$. That
approximately additional 1 MeV for $\left< V_{4}\right>$ is a big effect in
relation  to
$E_{t}$ itself, but only a small modification of the total potential energy.
 
The next point is to check the convergence in $j_{max}$ up to which the
3NF has to be taken into account (this corresponds to the $j_{max}$ of 
basis 1, the outer basis, in eq.~(\ref{4})). We see from Table \ref{tab2} that
the convergence 
of the triton binding energy with inclusion of the 3NF is slower than
without 3NF. This is displayed in detail in Table \ref{tab2a}. We see that
the contributions of the 3NF for a given
$j$, $(E_{t}|^{NN+3NF}_{j_{max}=j}-E_{t}|^{NN}_{j_{max}=j})
-(E_{t}|^{NN+3NF}_{j_{max}=j-1}-E_{t}|^{NN}_{j_{max}=j-1})$,
are larger than the contributions 
$E_{t}|^{NN}_{j_{max}=j}-E_{t}|^{NN}_{j_{max}=j-1}$
of the NN force (only for $j=6$ they are
both equally small). Further the contributions of the 3NF change their sign
with 
increasing $j$. For even $j$ the 3NF acts attractive and for odd $j$
repulsive. (Of course, the step to a higher $j$ also includes transition
potentials from lower $j$s to that specific higher $j$.)
This alternating sign in the contributions of the 3NF leads to the
fact that for odd $j$ the contribution of the 3NF to the triton binding energy
is partially cancelled by the NN force contribution. Therefore the total change
in the triton binding energy going from $j_{max}=4$ to $j_{max}=5$ is only 8
keV in this model. This 
is about 0.1\% of the binding energy and corresponds to our numerical
accuracy. Therefore we chose $j_{max}=5$ for the evaluation of
the triton binding energy including a 3NF.

A comparison to the work of the Pisa group using exactly the same force model 
shows very good agreement. The Pisa
group calculates the triton binding energy in configuration space using the
pair correlated hyperspherical harmonic basis approach \cite{kievsky94}. 
Thus their
method is mathematically and numerically totally independent from our
approach. The result given in \cite{kievsky94} is 8.484 MeV. Recently they
recalculated that number using more equations now and achieved 8.486 MeV
\cite{kievskypriv}. 
Both numbers are in excellent agreement to our most advanced
result of 8.486 MeV given in Table \ref{tab2}. 

The results shown in the following refer to $j_{max}=5$ both for the NN and
the 3NF. The inner $j_{max}$ is also restricted to that value 5.

Let us investigate now to the most modern, phase equivalent potentials AV18,
CD-Bonn, 
Nijm~93, I and II and Reid~93. All these potentials include charge
independence breaking (CIB) and on top AV18 and CD-Bonn also
charge symmetry breaking (CSB). In other
words, all these potentials are given in an np, pp and two also in an nn
version. The np and 
pp versions are obtained by fitting to the corresponding sets of NN phase
shifts, 
whereas the choices of the nn versions are nonuniform \cite{stokspriv2} and lead to
different behaviours, as will be shown below. Because of that and since only
little is known about the CSB in the NN force, we replace the nn by the pp
force; in other words, we neglect CSB.

On the other hand the CIB in the NN force is known rather well and we have to
take it 
into account. We do this according to \cite{witala89} \cite{witala91} 
by choosing
an effective $t$-matrix
\begin{equation}
\label{5}
t_{eff}=1/3\ t_{np}+2/3\ t_{pp}
\end{equation}
We neglect the total isospin $T=3/2$ admixture,
which is justified for the triton (and many but not all observables in nd
scattering \cite{report}). 

In Table \ref{tab1} we show the triton binding energies calculated for the
six potentials mentioned above using the effective NN $t$-matrix (\ref{5}).
We see the well known gap to the experimental number of 8.48 MeV. Only CD-Bonn
with its strong non-locality sticks out.
Just for the sake of information, we also present the results if we replace
$t_{pp}$ by $t_{nn}$ in eq.~(\ref{5}), which is possible for AV18 and
CD-Bonn. The results are shown in parenthesis and we find an effect of CSB of
about 100 and 60 keV, respectively.
This shift in energy is about what is needed to ``understand'' the mass
difference between $^{3}$H and $^{3}$He on top of electromagnetic effects and
n-p mass differences 
\cite{sasakawa86}
\nocite{ishikawa86}
-\cite{wu93}.
The larger effect for the AV18 potential is due to the fact that changes have
been introduced on an operator level in going from the pp to the nn system
\cite{stokspriv2}, whereas in CD-Bonn only the $^{1}S_{0}$ component of the NN
force has been changed \cite{machleidtpriv}. 
The theoretical binding energy for Nijm~II differs from the result shown in
\cite{glockle95a} \cite{glockle95b} 
because the old potential version has been used there; also we
increased now the accuracy in going to $j_{max}= 5 $. The difference to
AV18~(cd) shown in \cite{glockle95a} \cite{glockle95b} 
is again due to $j_{max}=5$, but more due to
the fact that in \cite{glockle95a} \cite{glockle95b} $t_{nn}$ 
instead of $t_{pp}$ has been used in
eq. (\ref{5}).
Also our numbers differ from the ones given in \cite{friar93}, since we allow
for CIB. We also included the Ruhrpot NN interaction \cite{ruhrpot}
\cite{eden96}  for reasons
explained below.

\begin{figure}

\input {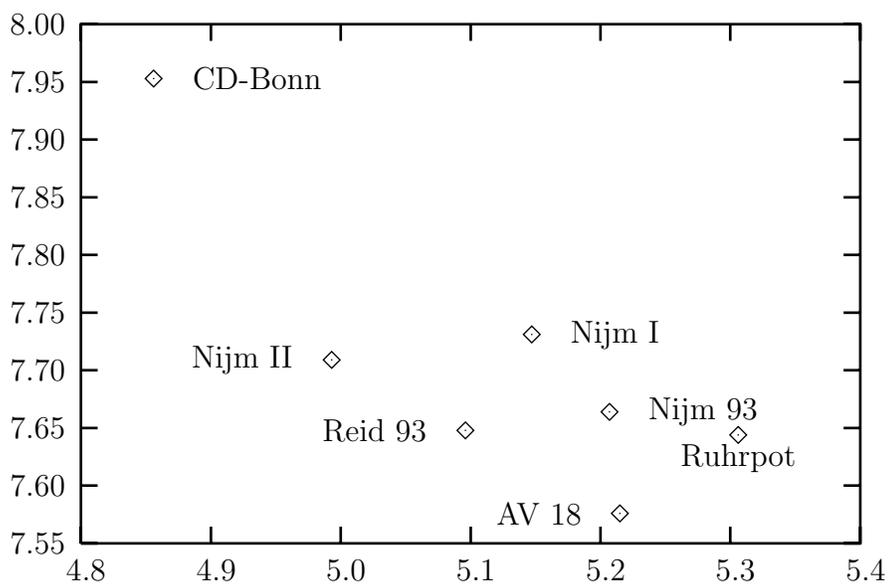}

\caption{\label{fig1} 
Triton binding energies versus $\Lambda$ according to Table \ref{tab1}.
}
\end{figure}

Now let us come to the main point, the adjustment of the triton binding energy
by choosing the appropriate cut-off parameter
$\Lambda$ in the strong form factors of the TM $\pi$-$\pi$ 3NF.
They are shown in Table \ref{tab1} together with the resulting binding
energies. The less accurate adjustment for the Ruhrpot interaction is
sufficient for the purposes discussed below.   Inspection of
Table \ref{tab1} reveals that the connection between $\Lambda$ and the triton
binding energy without 3NF is not linear as one might expect naively: for
example the 
two NN forces Reid~93 and Nijm~93 give nearly the same value for the triton
binding energy without 3NF, but their $\Lambda$s are quite different. This
fact demonstrates the subtle interplay of the 3NF with the various NN forces 
which can lead to unexpected results. Since the NN forces are phenomenological
and therefore no internal consistency exists to that 3NF, this is not
necessarily surprising and has been noticed before \cite{wiringa83}
\cite{glockle93}. We illustrate our findings
in Fig.~\ref{fig1}, were we plot the fitted $\Lambda$s
against the triton binding energy without 3NF.
Obviously the six potentials are divided into two groups: One group contains
Nijm~II, Reid~93 and AV18. For this group the connection between $\Lambda$ and
$E_{t}$ is very well linear. For the other group, CD-Bonn, Nijm~I and
Nijm~93 the connection is roughly linear.
Therefore the question arises, what distinguishes the potentials of
the one group from the potentials of the other group?

A natural guess is that the probability to find
the three nucleons in the triton at a certain distance from each other is
significantly different for the 
potentials of the two groups. A hint in that direction is the
NN correlation function in the triton, which is defined as
\begin{equation}
\label{6}
C(r)\equiv {1\over 3 } \ { 1 \over 4 \pi } \ \int d \hat r \left< \Psi
\right| \sum _{{i<j}} \ \delta(\vec r- \vec r_{{ij}}) \left| \Psi \right>
\end{equation}
where $r$ is the distance of two of the three nucleons and $ r_{ij}$ the
corresponding operator. $C$ provides the probability to find two nucleons at
a distance $r$. It is shown in Fig.~\ref{fig3} for the various NN potentials. 

 We see that there is indeed a significant difference between the
  potentials of the
 two groups mentioned above. For the potentials Nijm~II,Reid~93
 and AV18 $C$ is essentially zero at $r=0$, whereas the probabilities
  $C(r)$ are much less suppressed at short distances for the
 potentials CD-Bonn,Nijm~I,and Nijm~93. Apparently the different grouping of the potentials
  in Fig.~\ref{1} is related to the
 short range behaviour of $C(r)$ caused by those potentials.The
  potentials, which are strongly repulsive
 at short distances require a smaller strength factor $\Lambda $ in the 3NF
 to achieve the triton binding energy than  the weaker repulsive ones.For 
 instance Reid~93 and Nijm~93 give nearly the same triton binding energy(
 without 3NF) but require quite a different strength factor $\Lambda $. The one
 which allows two nucleons to come closer to each other, Nijm~93, needs
  a larger $\Lambda $. The
 corresponding remark applies to the pair Nijm~II and Nijm~I. CD-Bonn has
  no local
 , strongly repulsive partner to compare with , but the nearly linear
 correlation with $\Lambda $  for the three potentials Nijm~93,Nijm~I,
 and CD-Bonn shown in Fig.~\ref{fig1} has obviously to do with the increasingly
 weaker suppression of $C(r)$ at $r=0$. We also included a 7$^{th}$ potential,
 the Ruhrpot \cite{ruhrpot} \cite{eden96}, which is a meson theoretical interaction, but
 there the $\chi^{2}$ is not pushed to that accuracy as for the other
 potentials and it is fitted to a different set of NN phase shift parameters,
 namely the Arndt phases \cite{arndtphases}. 
Also the Ruhrpot model is provided only
 in a np version.  As seen from
 Fig.~\ref{fig3} $C(r)$ for that potential is also strongly suppressed near
 $r=0$ but nevertheless the $\Lambda$ is quite large. Its $C(r)$ behaves
 however differently from the others, since it rises very quickly to its
 maximum. We included that potential as an example for a qualitatively
 different behaviour of $C(r)$. Apparently that quick rise of $C(r)$ is more
 important than the strong suppression of $C(r)$ at very small $r \lapprox 0.2
 fm^{-1}$ and causes the large $\Lambda$. This demonstrates the subtlety of the interplay of properties of
 the NN forces and that 3NF. The detailed behaviour of the forces at short
 distances below about 1~fm is important.

 It is obvious that this observations can be further illustrated
  and understood by investigating
 the configuration space properties of that 3NF and the NN forces together
  with
 the behaviour of the 3N wavefunction. Since we work in momentum space
 this is not directly accessible to us and we leave that as a suggestion.

   In that context it is also of interest to see how $C(r)$ changes,
    once that 3NF has been included.
   Our results shown in Fig.~\ref{fig4} tell that the Cs do not
   change  qualitatively.
The $C$s increase in the maximum at $r \approx 1$~fm including that 3NF. This
is connected with the stronger decrease at larger $r$s. The change at very
short distances is nearly zero for the very repulsive NN potentials and
increases with decreasing repulsion. We also determined the probability to
find one nucleon at a certain distance from the centre of mass for the various
NN forces, with and without 3NF. The effect of that 3NF was to increase the
probability slightly for $r \ \lapprox \ 1\ fm$. Especially around 0.5~fm, where
the density without 3NF starts to flatten towards $r=0$ the density aquires a
small hump due to the 3NF. Our results are very similar to the one already
found in \cite{friar86}.

It will be interesting to repeat this sort of study for other 3NFs and to pin
down possible effects in inclusive and exclusive electron scattering. 

\acknowledgements{This work was supported by the Deutsche
Forschungsgemeinschaft and the Research Contract \# 41324878 (COSY-044)  of the
Forschungszentrum J\"ulich. The numerical
calculations have been performed on the T3E of the
H\"ochstleistungsrechenzentrum in J\"ulich, Germany. 
}

\bibliography{literatur}

\begin{table}
\begin{tabular} {c|c|c|c|c|c}
\multicolumn{6}{c}{AV14 only} \\
$j_{max}$&$E_{t}$ [MeV]&$\left< H_{0}\right>$ [MeV]
         &$\left< V\right>$ [MeV]
         &$\left< V_{4}\right>$ [MeV] &$\left< H\right>$ [MeV]\\
\hline
2&-7.577&45.177&-52.756&---&-7.579\\
3&-7.659&45.596&-53.257&---&-7.662\\
4&-7.674&45.654&-53.330&---&-7.676\\
5&-7.680&45.677&-53.360&---&-7.683\\
6&-7.682&45.680&-53.364&---&-7.684\\
\hline\hline
\multicolumn{6}{c}{AV14 + TM 3NF$^{\dagger}$}\\
$j_{max}$&$E_{t}$ [MeV]&$\left< H_{0}\right>$ [MeV]
         &$\left< V\right>$ [MeV]
         &$\left< V_{4}\right>$ [MeV] &$\left< H\right>$ [MeV]\\
\hline
2&-8.471&49.501&-56.553&-1.422&-8.474\\
3&-8.433&49.173&-56.332&-1.277&-8.436\\
4&-8.482&49.357&-56.525&-1.317&-8.485\\
5&-8.475&49.321&-56.498&-1.300&-8.478\\
\hline\hline
\multicolumn{6}{c}{AV14 + TM 3NF$^{\ddagger}$}\\
$j_{max}$&$E_{t}$ [MeV]&$\left< H_{0}\right>$ [MeV]
         &$\left< V\right>$ [MeV]
         &$\left< V_{4}\right>$ [MeV] &$\left< H\right>$ [MeV]\\
\hline
2&-8.478&49.516&-56.567&-1.430&-8.481\\
3&-8.440&49.185&-56.343&-1.285&-8.443\\
4&-8.490&49.370&-56.537&-1.325&-8.492\\
5&-8.482&49.332&-56.509&-1.308&-8.484\\
6&-8.486&49.340&-56.518&-1.310&-8.489\\
\end{tabular}

\caption{\label{tab2} 
Triton binding energy $E_{t}$ and energy expectation values $\left< H_{0}\right>$,
$\left< V\right>$,$\left< V_{4}\right>$ and $\left< H\right>$
for the AV14 NN potential only and together with the TM 3NF using 
$\Lambda=5.13\ m_{\pi}$.\hfill\break
$^{\dagger}$: 3NF calculated with $j_{max}=5$ for inner basis (see
text).\hfill\break 
$^{\ddagger}$: 3NF calculated with $j_{max}=6$ for inner basis (see text).
}
\end{table}

\begin{table}
\begin{tabular} {c@{\hspace{1cm}}|r@{\hspace{1cm}}|r@{\hspace{1cm}}}
&\multicolumn{1}{c}{AV14 only}\vline
&\multicolumn{1}{c}{AV14 + TM 3NF$^{\ddagger}$}\\
j&\multicolumn{1}{c}{$(E_{t}|^{NN}_{j_{max}=j}-E_{t}|^{NN}_{j_{max}=j-1})$ [keV]}\vline
&\multicolumn{1}{c}{$(E_{t}|^{NN+3NF}_{j_{max}=j}-E_{t}|^{NN}_{j_{max}=j})$}\\
&&\multicolumn{1}{c}{$-(E_{t}|^{NN+3NF}_{j_{max}=j-1}-E_{t}|^{NN}_{j_{max}=j-1})$ [keV]}\\ 
\hline
3&82&$781-901=-120$\\
4&15&$816-781=35$\\
5& 6&$802-816=-14$\\
6& 2&$804-802=2$\\
\end{tabular}
\caption{\label{tab2a} 
Contributions of the NN and the 3NF to the triton binding energy for a given
$j$ (see text).\hfill\break
$^{\ddagger}$: 3NF calculated with $j_{max}=6$ for inner basis.
}
\end{table}

\begin{table}
\begin{tabular} {c|l|l|l}
potential&$E_{t}$ [MeV]&$\Lambda /m_{\pi}$&$E_{t}$ [MeV]\\
&$t=1/3\ t_{np}+2/3\ t_{pp}$
&&$t=1/3\ t_{np}+2/3\ t_{pp}$\\
\hline
CD-Bonn&7.953 (8.014)&4.856&8.483\\
Nijm II&7.709&4.990&8.477\\
Reid 93&7.648&5.096&8.480\\
Nijm I &7.731&5.147&8.480\\
Nijm 93&7.664&5.207&8.480\\
AV18   &7.576 (7.685)&5.215&8.479\\
Ruhrpot &7.644&5.306&8.459\\
\end{tabular}
\caption{\label{tab1} 
Triton binding energies $E_{t}$ for various realistic NN potentials in
charge dependent calculations without $T=3/2$. The numbers in parenthesis
refer to np and nn forces. The adjusted cut-off parameters $\Lambda$ in the 3NF
with the resulting triton binding energies.
}
\end{table}

\pagebreak

\begin{figure}
\refstepcounter{figure}
\label{fig3} 

\input {corr.vergl.picture}

\input {corr.vergl.2.picture}

\begin{center}
FIG. \ref{fig3} Two-body correlation functions for the triton using various NN
potentials. Subfigure (b) shows an enlargement for small $r$ of subfigure (a).
\end{center}

\end{figure}

\pagebreak

\begin{figure}

{\scriptsize

\input{corr.cdbonn.picture}
\input{corr.nijm.1.picture}

\input{corr.nijm.2.picture}
\input{corr.nijm93.picture}

\input{corr.reid93.picture}
\input{corr.av18.picture}

\input{corr.ruhrpot.picture}

}

\begin{center}
\refstepcounter{figure}
\label{fig4}
\parbox[t]{1.5cm}{FIG. \ref{fig4}.\hfill}\parbox[t]{12.5cm}{Two-body 
correlation functions for the triton using various NN forces without (solid
line) and with (dashed line) the TM-3NF }  
\end{center}

\end{figure}

\end{document}